\documentclass{PoS}
\def\a{\bar{\alpha_s}}
\title{Jet fragmentation in a dense QCD medium}

\ShortTitle{Jet fragmentation in a dense QCD medium}

\author{\speaker{Paul Caucal}\\
        Institut de Physique Th\'{e}orique, Universit\'{e} Paris-Saclay, CNRS, CEA, F-91191, Gif-sur-Yvette, France\\
        E-mail: \email{paul.caucal@ipht.fr}}

\author{Edmond Iancu\\
       Institut de Physique Th\'{e}orique, Universit\'{e} Paris-Saclay, CNRS, CEA, F-91191, Gif-sur-Yvette, France\\
       E-mail: \email{edmond.iancu@ipht.fr}}
\author{Alfred H. Mueller\\
    Department of Physics, Columbia University, New York, NY 10027, USA\\
   E-mail: \email{amh@phys.columbia.edu} }
       
 \author{Gregory Soyez\\
       Institut de Physique Th\'{e}orique, Universit\'{e} Paris-Saclay, CNRS, CEA, F-91191, Gif-sur-Yvette, France\\
       E-mail: \email{gregory.soyez@ipht.fr}}

\abstract{
We study the fragmentation of a jet propagating in a dense quark-gluon plasma.  Using a leading,
double-logarithmic approximation in perturbative QCD, we compute for the first time the effects of
the medium on the vacuum-like emissions.  We show that, due to the scatterings off the plasma, the
in-medium parton showers differ from the vacuum ones in two crucial aspects:  their phase-space is
reduced and the first emission outside the medium can violate angular ordering.  We compute the
jet fragmentation function and find results in qualitative agreement with measurements at the LHC.
}

\FullConference{XIII Quark Confinement and the Hadron Spectrum - Confinement2018\\
		31 July - 6 August 2018\\
		Maynooth University, Ireland}

\begin{document}

\section{Introduction}

One of the main objectives of the experimental programs at RHIC and at the LHC is the characterisation of the quark-gluon plasma (QGP) produced in
ultrarelativistic heavy ion collisions.  An important class of observables used to study this dense form of QCD matter refers to the physics of
jet quenching, i.e.  the modifications of the properties of an energetic jet due to its interactions with the surrounding medium. 

For example the suppression of the jet cross-section in nucleus-nucleus collision w.r.t proton-proton is an indication of jet energy loss in the plasma \cite{Khachatryan:2016jfl,Aaboud:2018twu}. Most recently, 
substructure observables such that the jet fragmentation function \cite{Chatrchyan:2014ava,Aaboud:2018hpb}, the jet angular shape \cite{Chatrchyan:2013kwa} or the $z_g$ distribution \cite{Sirunyan:2017bsd} have also shed light on the evolution of a jet in the QGP. 
This proceeding focuses mainly on the jet fragmentation function which quantifies the particle distribution in energy inside jets.

From a theoretical point of view, high-$p_T$ jets are valuable because it is possible to rely on perturbative QCD to predict their properties. In that context, one of the effects of a dense 
weakly-coupled quark gluon plasma on virtual partons inside jets is to trigger medium-induced radiations because of the multiple collisions with the medium constituents. This can be
computed using the BDMPS-Z formalism \cite{Baier:1996kr,Zakharov:1996fv,Wiedemann:2000za}, recently generalised to include multiple medium-induced branchings 
\cite{Blaizot:2012fh,Blaizot:2013hx,Blaizot:2014ula,Kurkela:2014tla,Escobedo:2016jbm}. 
However it is also clear that the overall jet
structure should get built via the usual, ``vacuum-like'' bremsstrahlung through which a virtual parton evacuates its virtuality (until this becomes as small as the
hadronisation scale).

Thus, in order to construct the jet evolution in the presence of a medium, one needs to understand the interplay between these two mechanisms: vacuum-like emissions (VLEs) and medium 
induced radiations. Taken separately, these two mechanisms are by now rather well understood, but it appears as a challenge to construct a unified theoretical picture which consistently 
encompasses both sources of radiation.

In this proceeding based on the recent paper \cite{Caucal:2018dla}, we emphasize that the double-logarithmic approximation (DLA) of perturbative QCD is the first mandatory step toward a 
more advanced treatment of the evolution of jets in a QGP.  In the vacuum, DLA analysis exhibits the main physical ingredients of jets,
namely, the structure of intrajet parton cascades and the role of QCD coherence effects in soft gluon multiplication processes (hump-backed plateau) \cite{Dokshitzer:1991wu}.
With a dense medium, the same analysis can be done and  we show that within this approximation the time scales in the evolution factorize.

\section{Phase space for vacuum-like emissions with a dense medium}

In this section, we present the phase space available for vacuum-like emissions in the presence of a medium.

\subsection{Bremsstrahlung spectrum and BDMPS-Z spectrum}

\begin{figure}
\label{BremMed}
\begin{center} 
 \begin{tabular}{cc}
   \includegraphics[width=65mm]{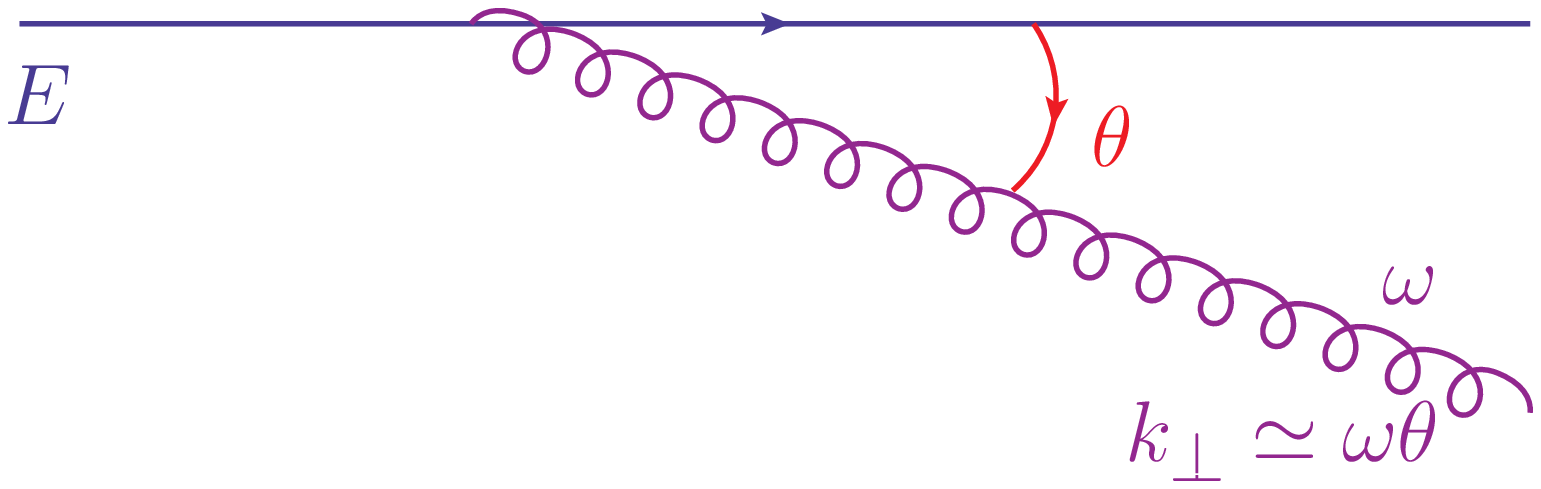}&
   \includegraphics[width=65mm]{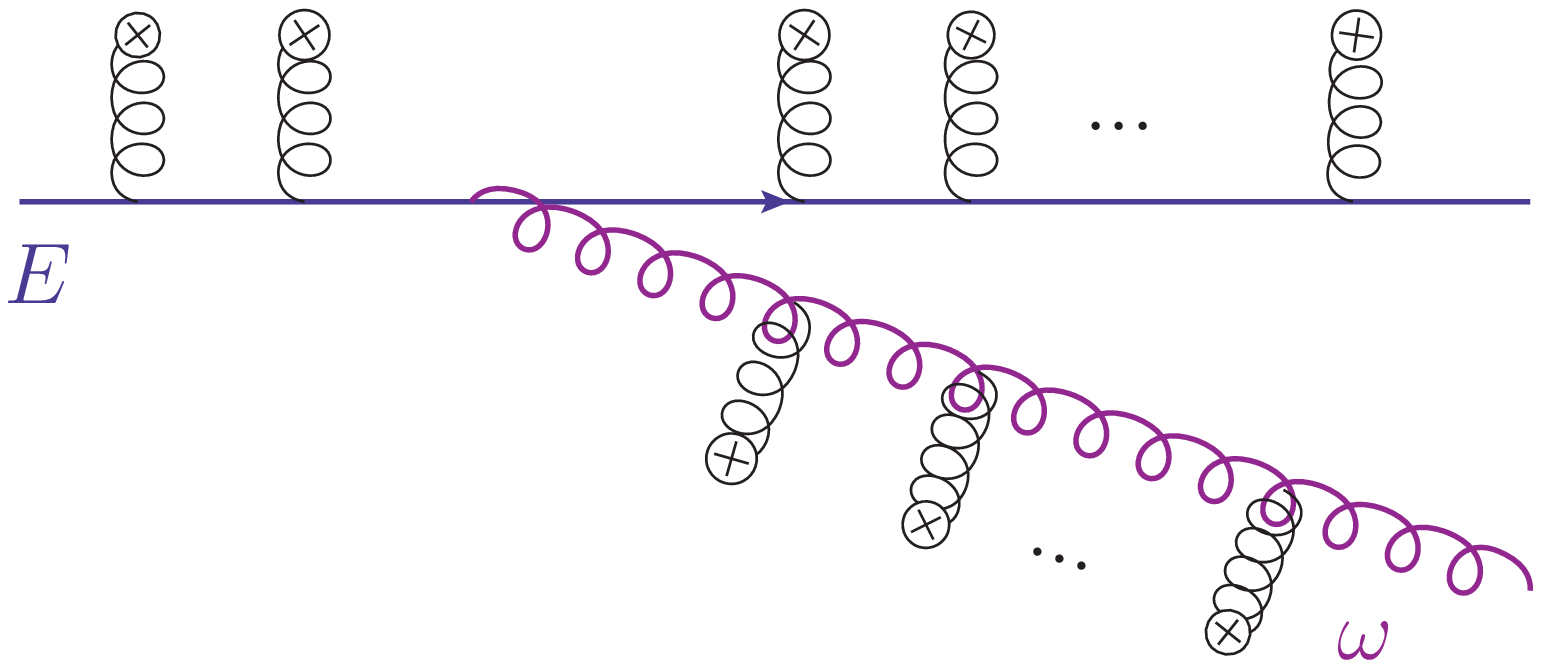}
   \\
\end{tabular}
\end{center}
\caption{Leading-order Feynman diagrams for the Bremsstrahlung (left) and medium-induced (right) processes. 
The medium is described by a fluctuating colored classical field whose vertices are represented by the symbol $\otimes$.}
\end{figure}

In the vacuum, the building block of a jet evolution is the Bremsstrahlung spectrum. A virtual parton produced by a 
hard scattering can radiate a gluon with a probability which is logarithmically enhanced for soft and collinear emissions. 
It is a specific feature of the 
Bremsstrahlung probability distribution $d\mathcal{P}_{\mathcal{B}}$ for the simple process represented figure \ref{BremMed}-left.
\[d\mathcal{P}_{\mathcal{B}}=\frac{\alpha_s C_R}{\pi}\frac{d\omega}{\omega}\frac{d\theta^2}{\theta^2}\]
This double-logarithmic enhancement of the emission probability is the essence of the double logarithmic approximation. 
In the computation of intrajet observables we resum to all orders only contributions of the form 
\[\alpha_s\log(\bar{\theta}^2/\theta_m^2)\log(E/\omega_m)\sim 1\]
for a given cut-off in energy $\omega_m$ and angle $\theta_m$,
with $\bar{\theta}$ is the opening angle of the jet and $E$ its energy. A loss of a single angular or energy logarithm is sub-leading at DLA.

On top of that, the dense weakly-coupled medium can trigger medium-induced radiations. The underlying hypothesis for such emissions to occur within the BDMPS-Z formalism 
is that the formation time of the medium-induced parton is much larger than the mean free path of the emitting particle. In the following, the quark-gluon plasma 
is characterized by only two parameters: the distance $L$ travelled by the jet inside the medium and the jet quenching parameter $\hat{q}$ related to the averaged 
transverse momentum acquired by multiple collisions during time $\Delta t$ by $\langle k_\perp^2\rangle=\hat{q}\Delta t$. In this framework, the probability distribution 
$d\mathcal{P}_{BDMPS-Z}$ for the process represented figure \ref{BremMed} is well approximated by the formula \cite{Salgado:2003gb,Wiedemann:2000tf} 
\[d\mathcal{P}_{BDMPS-Z}\simeq \frac{\alpha_sN_c}{\pi}L \sqrt{\frac{\hat{q}}{\omega^3}} d\omega\]

With respect to the Bremsstrahlung spectrum, the obvious property of the BDMPS-Z spectrum is that collinear and soft radiations are not \textit{logarithmically} enhanced. This will 
have important consequences since working at the double-log accuracy enables  to simply ignore such emissions for the intrajet activity.

\subsection{Vacuum-like emissions inside and outside the medium}

However, as we show in this subsection, medium induced radiations provide a natural constraint on the phase space available for one emission inside the medium.

Indeed, considering first an emission which occurs inside the medium, its
formation time $t_f$ is then smaller than $L$. $t_f$ is determined by the uncertainty principle, namely the condition that the transverse separation 
$\Delta r \sim \theta t_f$ between the gluon and its parent parton at the time of emission to be as large as the gluon transverse wavelength
$2/k_\perp$, with $k_\perp\simeq \omega\theta$ its transverse momentum w.r.t its parent. This argument applies to both vacuum-like
and medium-induced emissions and implies $t_f\simeq 2\omega/k_\perp^2\simeq 2/(\omega\theta^2)$.  Then, gluons emitted inside the medium
have a minimum $k_\perp$ set by the momentum acquired via multiple collisions during its formation, $k_f^2=\hat{q}t_f$. Gluons produced inside the medium 
with a transverse momentum smaller than $\hat{q}t_f$ cannot exist. This translates into an upper limit $t_f\le\sqrt{2\omega/\hat{q}}$
on the formation time of any gluon inside the medium. That said, medium induced gluons for which $k_\perp\simeq k_f$ are excluded because the emission probability 
is not enhanced by double logarithms. Consequently, at DLA the only contribution to the intrajet activity inside the medium comes from 
vacuum-like (Bremsstrahlung) emissions with $k_\perp \ge k_f$, inequality which becomes strong at DLA: $k_\perp \gg k_f$ or equivalently 
$t_f\ll \sqrt{2\omega/\hat{q}}$ meaning that VLEs occur much faster than medium-induced radiations with the same energy.

This is the condition for an emission \textit{inside} the medium to be vacuum-like. For a medium with a fixed finite length, 
an emission can also occur directly outside if its formation time is larger than $L$: $t_f\ge L$.

\subsection{Lund diagram for one vacuum-like emission}

\begin{figure}
 \label{phasespace}
 \begin{center}
  \begin{tabular}{cc}
 \includegraphics[width=71mm]{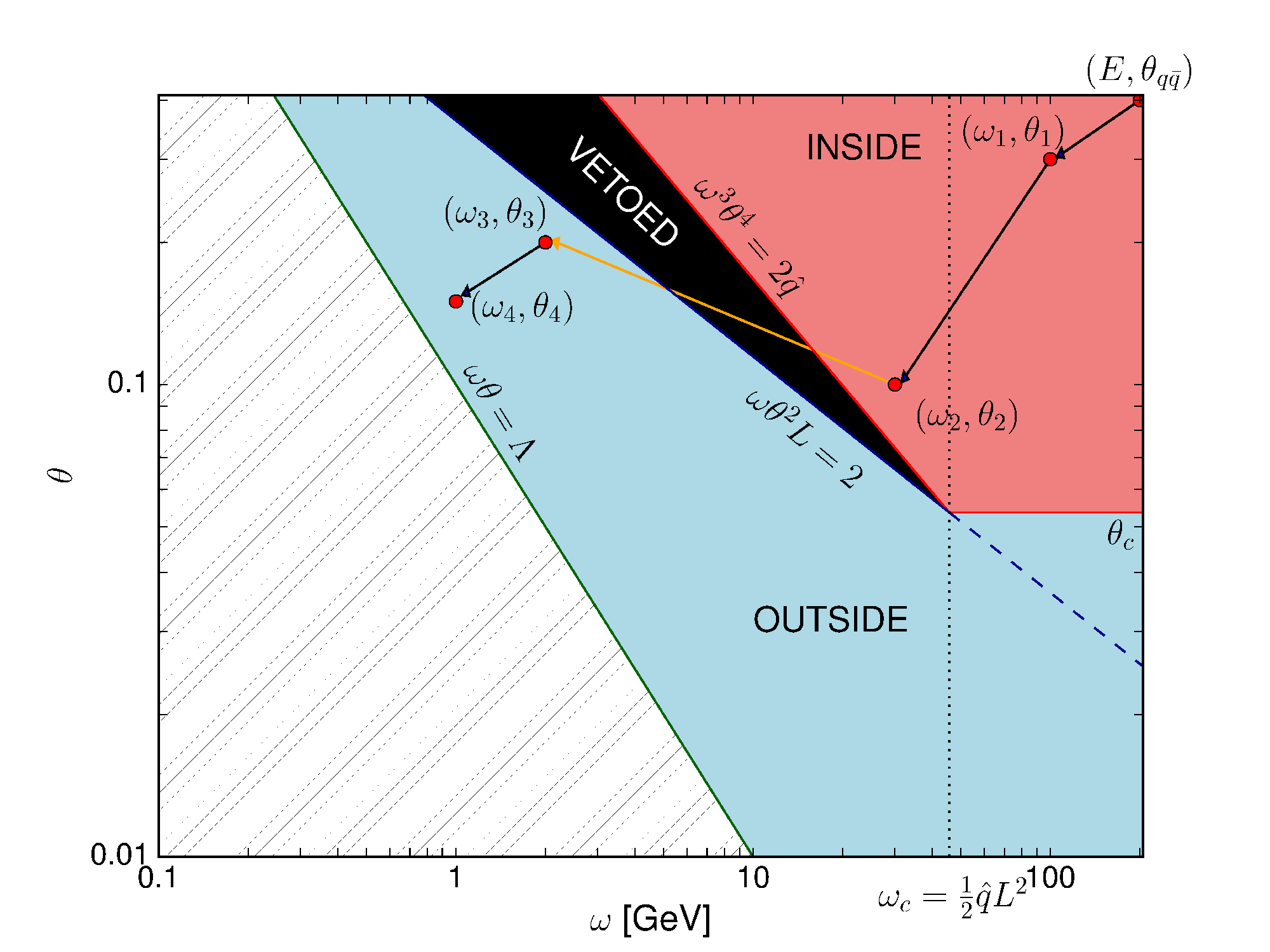}&
 \includegraphics[width=71mm]{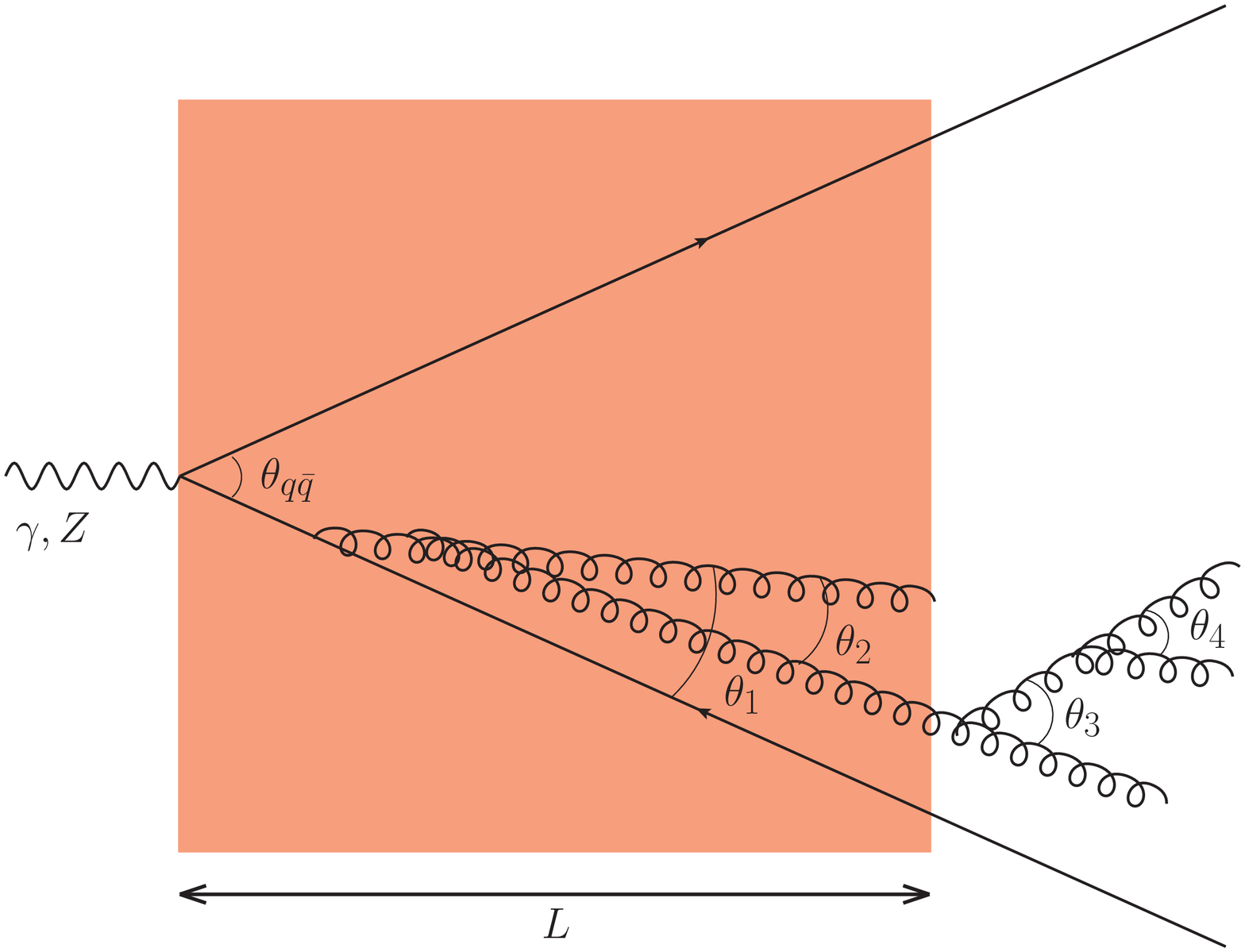}
 \end{tabular}
 \end{center}
 \caption{(Left) Phase space for one emission in a jet of energy $E=200$ GeV and opening angle $\bar{\theta}=0.4$ in the presence of a medium. 
The red region corresponds to the condition $t_f\le\sqrt{2\omega/\hat{q}}$ and $\theta\ge\theta_c$ and the blue 
 region $t_f\ge L$. In the hatched region, partons have hadronised and the perturbative QCD regime is not valid anymore. (Right) Typical cascade which may occur
 with a medium and not in the vacuum. The position of corresponding successive emissions in the phase space are represented on the left. The first two 
 emissions occur inside the medium and the last two outside. Angular ordering is violated by the third emission, but preserved by all the others.}
\end{figure}

These two conditions for VLEs, either inside with $t_f\ll \sqrt{2\omega/\hat{q}}$ or outside with $t_f\ge L$ can be written 
in terms of their energies $\omega$ and angles w.r.t the emitter $\theta$:
\[t_f\ll \sqrt{2\omega/\hat{q}}\Longleftrightarrow\omega\gg\omega_0(\theta)=(2\hat{q}/\theta^4)^{1/3}\textnormal{  and  } t_f\ge L\Longleftrightarrow\omega\le\omega_L(\theta)=2/(\theta^2L)\]
It is enlightening to represent these conditions on a diagram with $\omega$ and $\theta$ as axis. Compared to the vacuum case for which the phase space is 
only restricted by the hadronisation line $k_\perp\simeq\omega\theta=\Lambda$, the effect of the medium on the phase space available for VLEs is the presence of 
a vetoed region \cite{Caucal:2018dla} where there is no VLE permitted as shown figure \ref{phasespace}.

The critical energy $\omega_c=1/2\hat{q}L^2$ such that $\sqrt{2\omega/\hat{q}}=L$ also appears figure \ref{phasespace}. 
Emissions with larger energies $\omega\ge\omega_c$ behave exactly as
in the vacuum: their emission angle can be arbitrarily small and their formation time can be larger than $L$.
We shall assume that $E\ge \omega_c$, which is indeed the case for the high energy ($E\ge100$ GeV) jets at the LHC.

\section{Construction of the parton shower with a dense medium}

In the previous section, the phase space for one VLE at DLA has been discussed. Here, we explain how to iterate such emissions and how the medium affects 
the standard angular ordered vacuum cascade.

\subsection{Decoherence}

In the vacuum quantum color coherence implies angular ordering: a jet is described in terms of a
classical shower picture with successive angular ordered Bremsstrahlung emissions forming the Markov
chains of independent elementary radiation events \cite{Dokshitzer:1991wu}. 
However a $q\bar{q}$ antenna with opening angle $\theta_{q\bar{q}}$
propagating through a dense QGP can lose its coherence via rescattering off the medium: the quark and the
antiquark suffer independent color rotations, hence the probability that the antenna remains in a color singlet
state decreases with time \cite{MehtarTani:2010ma,CasalderreySolana:2011rz,CasalderreySolana:2012ef}. The two legs of the antenna start behaving like independent color sources after a
time $t_{coh}(\theta_{q\bar{q}})\equiv(4/\hat{q}\theta_{q\bar{q}}^2)^{1/3}$ \cite{CasalderreySolana:2011rz}. Consequently, angular ordering could in principle be violated inside the medium.

\subsection{Iteration of VLEs inside the medium}
It is conceptually simpler to consider a jet initiated by a quark-antiquark antenna in a color singlet state
with opening angle $\theta_{q\bar{q}}\le1$, e.g. produced by the decay of a boosted W/Z boson or a virtual photon. (For a
generic jet which is produced by a parton, the role of $\theta_{q\bar{q}}$ would be played by the jet radius $\bar{\theta}$.) The quark and
the antiquark are assumed to have equal energies $E/2$. Also, the antenna is assumed to be produced directly inside the
medium and to cross the medium along a distance L. Finally, for simplicity we shall work in the limit of a large
number of colors $N_c\gg 1$, where a gluon emission can be pictured as the splitting of one dipole into two.

Now one can show that decoherence has no consequences on the development of the shower \textit{inside} the medium. 
Indeed, decoherence is impossible for VLEs inside the medium. Consider the $i^{th}$ antenna in the evolution: 
its energy is $\omega_i$ and its opening angle is $\theta_i$. It has been emitted by the antenna $i-1$. 
By definition $\theta_{0}=\theta_{q\bar{q}}$ and $\omega_0=E$. Then it is easy to check that 
 \[t_f(\omega_i,\theta_i)\ge t_{coh}(\theta_{i-1})\textnormal{  and  }\theta_i\ge\theta_{i-1} \Longrightarrow t_f(\omega_i,\theta_i)\ge \sqrt{2\omega_i/\hat{q}}\]
 so that an incoherent large angle emission is necessarily inside the vetoed region and therefore is not allowed at DLA. 
 
Hence, at DLA, successive in-medium vacuum-like emissions are strongly 
ordered both in energy and angle \cite{Caucal:2018dla}.
 
\subsection{First emission outside the medium}

Actually, the argument presented above does not apply if the antenna $i-1$ is the \textit{last} antenna inside the medium. Indeed, in that case, the condition for 
the $i^{th}$ antenna to be vacuum-like is not $t_f(\omega_i,\theta_i)\ll \sqrt{\frac{2\omega_i}{\hat{q}}}$ but $t_f(\omega_i,\theta_i)\ge L$. 
Therefore the good criterion to discuss the coherence property of the antenna $i-1$ is the condition $t_{coh}(\theta_{i-1})=L$ since the next antenna will 
require a time larger than $L$ to be emitted. The critical angle $\theta_c=2/\sqrt{\hat{q}L^3}$ satisfies the equality $t_{coh}(\theta_c)=L$. Thus, if $\theta_{i-1}\le\theta_c$, the coherence time is also larger than $L$ and angular ordering is preserved. On the other hand, if $\theta_{i-1}\ge\theta_c$ 
the antenna has lost its coherence during the formation time of the next antenna so there is no constraint on the angle $\theta_i$ of the next antenna. The introduction 
of this angle $\theta_c$ leads to a slight modification of the in-medium region in the Lund diagram: this region is now defined by $t_f(\omega)\le \sqrt{\frac{2\omega}{\hat{q}}}$ and 
$\theta\ge\theta_c$ so that inside this region, cascades are always angular ordered.

To sum up, decoherence induced by the medium has only one effect on the parton shower at DLA: one violation of angular ordering by the first emission
outside the medium is permitted \cite{Caucal:2018dla} (a similar idea appears in \cite{Mehtar-Tani:2014yea}). Nevertheless this effect will have striking consequences on the shape of the fragmentation function.

\section{Calculation of the fragmentation function}

In this section, the fragmentation function of a jet at parton level is calculated according to the principles that we have just established.

\subsection{DLA results}

The basic quantity required to compute the fragmentation in our picture is the double differential inclusive probability distribution to find a parton with 
energy $\omega$ and angle $\theta^2$ (w.r.t its emitter) inside the jet of energy $E$ and angle $\theta_{q\bar{q}}$.
\[T(\omega,\theta^2|E,\theta_{q\bar{q}}^2)\equiv \omega\theta^2\frac{dN}{d\omega d\theta^2}\]
At DLA in the vacuum, this function satisfies the simple following master equation \cite{Dokshitzer:1991wu}
\[T_{vac}(\omega,\theta^2\mid E, \theta_{q\bar{q}}^2) = \a
 +\a\int_{\theta^2}^{\theta_{q\bar{q}}^2}\frac{d\theta_1^2}{\theta_1^2}\int_{\omega/E}^{1}\frac{dz_1}{z_1}T_{vac}(\omega,\theta^2\mid z_1E, \theta_1^2)\]
 with $\a=\alpha_s N_c/\pi$. The solution to this equation is \[T_{vac}(\omega,\theta^2\mid E, \theta_{q\bar{q}}^2)=\a I_0\big(2\sqrt{\a\log(E/\omega)\log(\theta_{q\bar{q}}^2/\theta^2)}\big)\]
 with $I_0$ the modified Bessel function of rank 0.
 
In the in-medium region of the phase space or for $\omega\ge\omega_c$, since nothing differs from the vacuum case, 
the corresponding function with a medium $T(\omega,\theta^2)$ 
is identical to $T_{vac}(\omega,\theta^2)$. However, in the out-of-the-medium region, $T$ is different because one must take into account 
``jumps'' over the vetoed region with a possible violation of angular ordering. Mathematically, this can be done by convolutions of the function $T_{vac}$ over 
the disjoint in and out regions of the phase space \cite{Caucal:2018dla}.

Once $T(\omega,\theta^2|E,\theta_{q\bar{q}}^2)$ is known for every $\omega$ and $\theta^2$, the fragmentation function $D(\omega|E,\theta_{q\bar{q}}^2)$ at the parton level is obtained by integrating 
out the unmeasured angle $\theta^2$ between the $k_\perp=\Lambda$ cut-off and $\theta_{q\bar{q}}$.
\[D(\omega|E,\theta_{q\bar{q}}^2)\equiv \omega\frac{dN}{d\omega}=\int_{\Lambda^2/\omega^2}^{\theta_{q\bar{q}}^2}\frac{d\theta^2}{\theta^2}T(\omega,\theta^2|E,\theta_{q\bar{q}}^2)\]

\begin{figure}
 \label{distrib}
 \begin{center}
  \begin{tabular}{cc}
 \includegraphics[width=71mm]{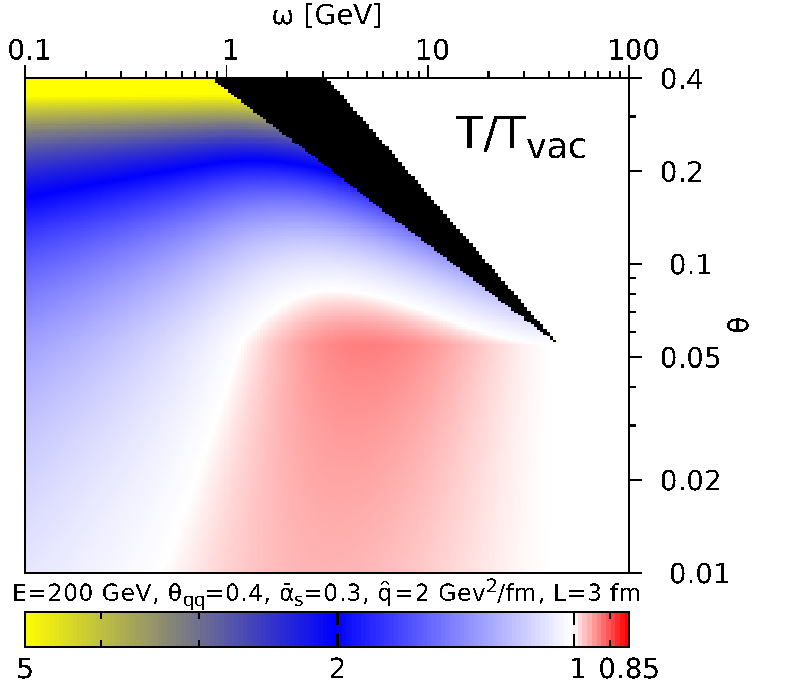}&
 \includegraphics[width=71mm]{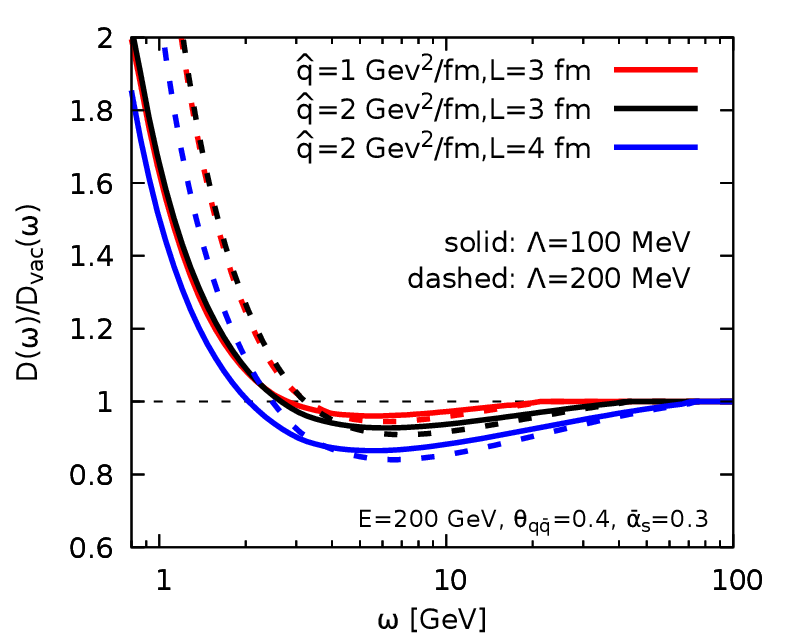}
 \end{tabular}
 \end{center}
 \caption{(Left) The ratio $T(\omega,\theta^2)/T_{vac}(\omega,\theta^2)$ between the two-dimensional gluon distributions in the medium and respectively the vacuum, 
 both computed to DLA and for the values of the free parameters $E$, $\theta_{q\bar{q}}$, $\a$, $\hat{q}$ and $L$ shown in the
figure. (Right) The ratio $D(\omega)/D_{vac}(\omega)$ between the fragmentation functions in the medium and respectively the vacuum,
for different choices for the medium parameters $\hat{q}$ and $L$ and the hadronisation scale $\Lambda$.}
\end{figure}

The results from \cite{Caucal:2018dla} are shown figure \ref{distrib}. The left figure refers to the two-dimensional gluon distribution $T(\omega,\theta^2|E,\theta_{q\bar{q}}^2)$: 
we more precisely show the ratio $T(\omega,\theta^2)/T_{vac}(\omega,\theta^2)$ between the distribution generated in 
the presence of the
medium and that in the vacuum. This ratio is $1$ for all the points either inside the medium or with $\omega\ge\omega_c$ as expected.
However, one sees significant deviations from unity for points outside the medium with energies $\omega\le\omega_c$:
for intermediate values of $\omega$ and relatively small angles $\theta\simeq0.1\theta_{q\bar{q}}$, one sees a small but significant
suppression compared to the vacuum (up to 15\%). For smaller energies and larger angles, $\theta\ge 0.2$, one
rather sees a strong enhancement, owing to emissions violating angular ordering.

The right plot in figure \ref{distrib} shows the ratio $D(\omega)/D_{vac}(\omega)$. One sees a slight suppression
(relative to vacuum) at intermediate energies, roughly from 2 GeV up to $\omega_c$, and a substantial enhancement
at lower energies $\omega\lesssim2$ GeV. This enhancement is attributed to small-angle emissions inside the medium,
radiating at larger angles outside the medium due to the lack of angular ordering. These results are in
qualitative agreement with the respective LHC measurements for the most central PbPb collisions \cite{Chatrchyan:2014ava,Aaboud:2018hpb}.

\subsection{Running coupling and single-log corrections}

Previous calculations assumed a strong ordering both in energy and angle and a fixed
coupling $\alpha_s$ (an assumption valid in the double-log approximation). It is natural to ask whether our results for the fragmentation
function are robust enough even if we relax one of these hypothesis.
For instance, one can relax the fixed coupling approximation or take into account single-log corrections.

\begin{figure}
 \label{NDLA}
 \begin{center}
  \begin{tabular}{cc}
 \includegraphics[width=71mm]{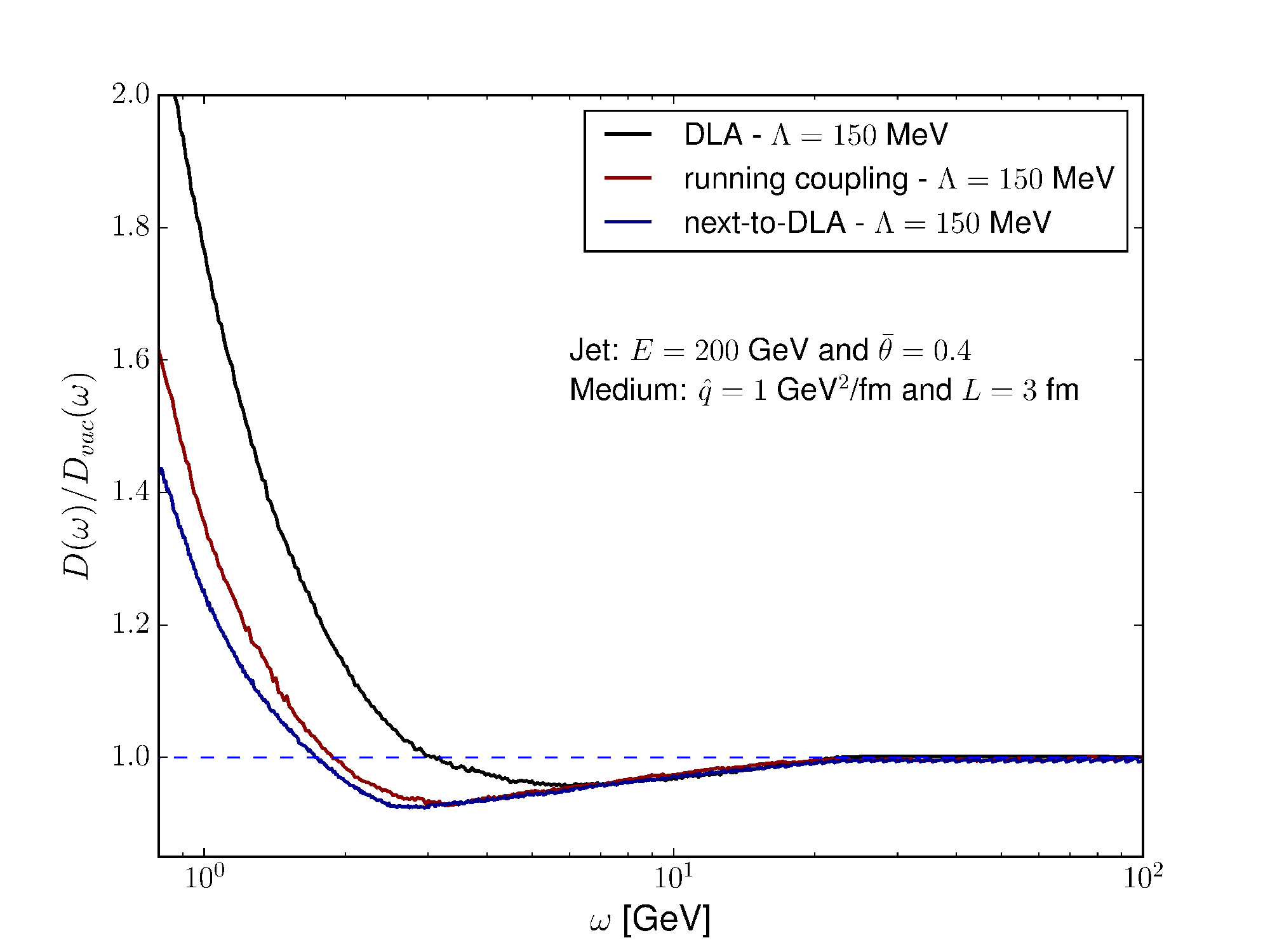}&
 \includegraphics[width=71mm]{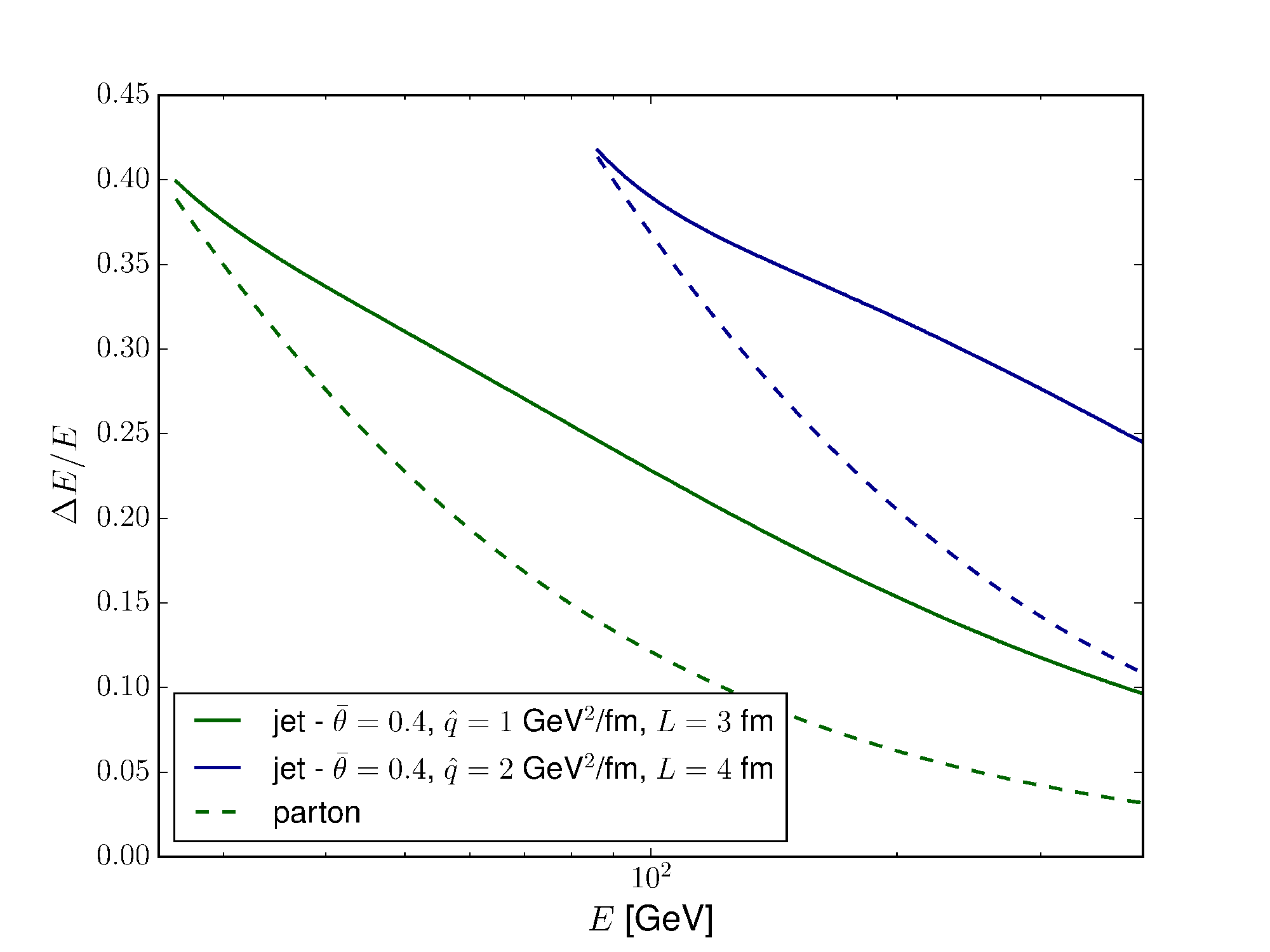}
 \end{tabular}
 \end{center}
 \caption{(Left) The ratio $D(\omega)/D_{vac}(\omega)$ between the fragmentation functions in the medium and respectively the vacuum for different approximations 
 in perturbative QCD: DLA, running coupling and single-log corrections from the full splitting function. 
 (Right) The relative energy loss by a jet $\Delta E/E$ due to medium induced radiations as a function of the jet energy $E$. The dashed lines represent the same quantity 
 for a single parton propagating through the medium.}
 \end{figure}

This can be done in a straightforward way in this formalism. If only gluons are considered, 
the general master equation for $T_{vac}$ is indeed \cite{DOKSHITZER1993137,Lupia1998}
\[T_{vac}(\omega,\theta^2\mid E, \bar{\theta}^2) = \a \frac{\omega}{E} P_{gg}(\omega/E)
 +\int_{\theta^2}^{\bar{\theta}^2}\frac{d\theta_1^2}{\theta_1^2}\int_{\omega/E}^{1}dz_1\a(z_1^2E^2\theta_1^2)P_{gg}(z_1)T_{vac}(\omega,\theta^2\mid z_1E, \theta_1^2)\]
with $P_{gg}(z)=(1-z)\big[z(1-z)+\frac{1-z}{z}+\frac{z}{1-z}\big]$.
 Then the gluon distribution with a medium $T$ is calculated by convolutions using the new solution $T_{vac}$ of this equation. 
 Numerical results are shown figure \ref{NDLA}-left for three cases:
 \begin{itemize}
  \item the double-log approximation corresponds to $\a=cste$ and $P_{gg}(z)\simeq1/z$.
  \item the running of the coupling: $\a(\omega^2\theta^2)=\frac{1}{\bar{b}}\frac{1}{\log(\omega^2\theta^2/\Lambda^2)}$, $\bar{b}=11/12$ and $P_{gg}(z)\simeq1/z$.
  \item the next-to-double-log approximation: the running coupling and the finite part of the splitting function are taken into account,
  namely $P_{gg}(z)\simeq\frac{1}{z}+\int_{0}^{1}dz\big(P_{gg}(z)-\frac{1}{z}\big)=\frac{1}{z}-\frac{11}{12}$ \cite{Lupia1998}.
 \end{itemize}
As expected, DLA overestimates the number of soft gluons inside the jet but the enhancement at small energies is still significant if we include 
the effects of the running coupling and the full splitting function. This shows the robustness of decoherence to generate soft gluons inside 
jets in the presence of a medium.
 
 \subsection{Jet energy loss}
 
Even if the next-to-double-log approximation does not conserve the energy since the symmetry $z\Leftrightarrow1-z$ is broken, one can try to estimate the energy loss for a jet compared to the energy lost by 
a single parton propagating through the medium.

First of all, at DLA energy loss is negligible for any parton of the cascade inside the medium, except for the \textit{last} one
which will propagate through the medium over a distance of order L. Indeed, the maximal energy loss $\omega_{loss}(t_f(\omega_i,\theta_i))$ 
by the $i^{th}$ parton of the in-medium cascade during 
its formation time is of the order $\hat{q}t_f^2/2$, the energy of the hardest medium induced
emission that can develop during $t_f$ \cite{Baier:1996kr,Zakharov:1996fv}. By the inequality $t_{f}(\omega_i,\theta_i^2)\ll \sqrt{2\omega_i/\hat{q}}$, one 
finds that $\omega_{loss}(t_f)\ll\omega_i$.

For an estimation of the energy loss by a jet as a global physical object, the DLA picture has an interesting
property: the partons produced inside the medium via DLA cascades act as new sources which will lose energy via
medium induced processes. The energy loss by a single source of energy $\omega$ propagating over a distance $L$  is estimated from medium
induced radiation of BDMPS-Z type iterated using
energy-conserving splitting vertex (essential at this stage since we focus on energy) \cite{Blaizot:2013hx}. It is a good approximation to estimate the energy lost
at large angles as the energy accumulated by the very
soft quanta which thermalize, according to the formula \cite{Fister2015}
\[\mathcal{E}(\omega,L)\simeq \omega(1-e^{-2\pi\omega_{br}/\omega})\]
with $\omega_{br}=\alpha_s^2\omega_c$ the
scale below which multiple medium-induced branchings become important. A source loses typically an energy $\textnormal{min}(\omega,\omega_{br})$. The total 
energy lost by a jet from an initial parton with energy $E$ is 
\[\Delta E=\mathcal{E}(E,L)+\int_{\theta_c^2}^{\theta_{q\bar{q}}^2}\frac{d\theta_1^2}{\theta_1^2}\int_{\omega_{0}(\theta_1)}^{E}\frac{d\omega_1}{\omega_1}T(\omega_1,\theta_1^2|E,\theta_{q\bar{q}}^2)\mathcal{E}(\omega_1,L)\]

It would be necessary to impose energy conservation in all vertices of the cascade to have a good quantitative estimation of the energy loss,
but even in the result figure \ref{NDLA}-right,
one sees that jet energy loss decreases with its initial energy more slowly than the parton energy loss. 
This is a first step toward a precise understanding of the $R_{AA}$ ratio for jets in the data \cite{Khachatryan:2016jfl,Aaboud:2018twu} since the vacuum inclusive jet cross-section is not included in this calculation.
 
 \section{Conclusion}
 
In this proceeding, we showed that vacuum-like emissions inside the medium can be factorized from the medium-induced radiations within
the double-log approximation. This approximation is fine for intrajet multiplicity at small energy but it is not accurate enough for experimental observables 
relying on energy because energy is not exactly conserved through the jet evolution.

However, DLA gives a new insight on the development of a jet in the quark-gluon plasma. Since based on a probabilistic picture, this approach is suitable for Monte-Carlo implementations, which
would allow to go beyond the present approximations and reach more quantitative predictions.

\vspace{0.5cm}
\textbf{Acknowledgements.}  The work of P.C, E.I. and G.S. is supported in
part  by  the  Agence  Nationale  de  la  Recherche  project ANR-16-CE31-0019-01.
The  work  of  A.H.M.  is  supported in part by the U.S. Department of Energy Grant
\# DE-FG02-92ER40699.

\bibliographystyle{JHEP}
\bibliography{biblio}

\end{document}